\documentclass[reprint,pra,
notitlepage,longbibliography]{revtex4-1}

\usepackage{graphicx} 
\graphicspath{{./figures/}}
\usepackage{float}
\usepackage{array} 
\usepackage{paralist} 
\usepackage{subfig}
\usepackage{fancyhdr} 
\usepackage{threeparttablex}
\usepackage{amsmath}
\usepackage{mathtools}
\usepackage{tikz-inet}	
	\usetikzlibrary{graphs,trees,shapes,snakes,arrows,circuits.logic.US}
\usetikzlibrary{backgrounds,fit,decorations.pathreplacing} 
\tikzstyle{branch}=[fill,shape=circle,minimum size=3pt,inner sep=0pt]
\usepackage{xcolor}

\usepackage[qm]{qcircuit}
\newcommand{\ket}[1]{\ensuremath{\left|#1\right\rangle}} 
\newcommand{\bra}[1]{\ensuremath{\left\langle#1\right|}} 

\newcommand{\alert}[1]{\color{red}#1} 
\begin{document}
\title{An Ancilla Based Quantum Simulation Framework for Non-Unitary Matrices} 
 \author{Ammar~Daskin}\email{also known as Anmer Daskin. email: adaskin25@gmail.com}
\affiliation{Department of Computer Engineering, Istanbul Medeniyet University, Kadikoy, Istanbul, Turkey}
 \author{Sabre~Kais}
\affiliation{Department of Chemistry, Department of Physics and Birck Nanotechnology Center, Purdue University, West Lafayette, IN, USA}
\affiliation{Qatar Environment and Energy Research Institute, HBKU, Doha, Qatar} 
\begin{abstract}
The success probability in an ancilla based circuit generally decreases exponentially in the number of qubits consisted in the ancilla. Although the probability can be amplified through the amplitude amplification process, the input dependence of the amplitude amplification makes difficult to sequentially combine two or more ancilla based circuits.
A new version of the amplitude amplification known as the oblivious amplitude amplification runs independently of the input to the system register. This allows us to sequentially combine two or more ancilla based circuits. 
However, this type of the amplification only works when the considered system is unitary or non-unitary but somehow close to a unitary.

In this paper, we present a general framework to simulate non-unitary processes on ancilla based quantum circuits in which the success probability is maximized by using the oblivious amplitude amplification. 
In particular, we show how to extend a non-unitary 
matrix to an almost unitary matrix. We then employ the extended matrix by using an ancilla based circuit design along with the oblivious amplitude amplification.
Measuring the distance of the produced matrix to the closest unitary  matrix, 
a lower bound for the fidelity of the final state obtained from the oblivious amplitude amplification process is presented. 
Numerical simulations  for random matrices of different sizes show that independent of the system size, the final amplified probabilities are generally around  0.75 and the fidelity of the final state is mostly high and around 0.95. 
Furthermore, we discuss the complexity analysis and show that combining two such ancilla based circuits, a matrix product can be implemented. 
This may lead us to efficiently implement matrix functions represented as infinite matrix products on quantum computers.
\end{abstract}

\maketitle
\section{Introduction and Background}
Quantum computers function by using the unitary time evolution operators, i.e. quantum gates, described in the formalism of the standard quantum mechanics. 
  Although  non-unitary dynamics can be emulated in the scope of unitary dynamics by using ancilla or measurement based quantum circuits (e.g., \cite{Wang2010Measurement, Terashima2005nonunitary,Daskin2014universal}), the success probability in these circuits diminish substantially with the size of the ancilla register. 
In most cases, the small success probability can be remedied by using  the standard amplitude amplification \cite{Grover1998,Mosca1998quantum, Brassard2002}. 
However, the input dependence of the amplitude amplification makes difficult to sequentially combine two or more ancilla based circuits.
In certain types of ancilla based quantum circuits, it is shown that the amplitude amplification can be done on the ancilla system independently of the input to the system register. This type of the amplification process is named as the oblivious amplitude amplification \cite{Paetznick2014repeat, Berry2014exponential}. In different contexts, it is also named as repeat untill success algorithm \cite{Lim2005RUS,Lim2006RUS} (please see Ref. \cite{Kothari2014efficient} for a broad overview of the topic).
A slightly modified version of the oblivious amplitude amplification is proven to be working also for matrices which are almost unitary: i.e., a non-unitary matrix but somehow close to a unitary matrix \cite{Berry2015Taylor}. 

A Hermitian matrix, $A$, can be used along with $\sqrt{I-A}$ ($I$ is an identity matrix) to form a unitary matrix which can then be processed on an ancilla based quantum circuit. 
However, finding the square root of a matrix is numerically a difficult problem when the size of the matrix is large. 
 In this paper, we show how to approximate this unitary matrix formation in a way that the produced matrix is almost unitary.
In addition, we show that this type of approximate matrices can be processed on the ancilla based circuit where the success probability is maximized by the oblivious amplitude amplification. 
We also present a lower bound for the fidelity of the final state obtained from the amplification process by measuring the distance of the approximate matrix to the closest unitary matrix.
Numerical simulations are done for random matrices of different sizes and it is found that independent of the system size, the final amplified probabilities are generally around  0.75 and the fidelity of the final state is mostly high and around 0.95. This makes possible to implement a product of non-unitary matrices  on quantum computers by sequentially combining two ancilla based circuits. Therefore, it may lead us to efficiently implement matrix functions represented as infinite matrix products.

This paper is organized as follows: In the following two subsections,    the ancilla based circuits considered in this paper and the amplitude amplification are introduced for the unfamiliar readers. In the next section, first, it is described how to complete a non-unitary matrix to a unitary matrix and then how to approximate it. Then, the numerical examples and the circuit design used in the examples are shown. In Sec.\ref{SecDiscus}, the algorithmic complexity of the whole framework is analysed and its possible use in the matrix product and matrix functions is discussed. Finally, the paper is concluded with the summary of the results.  
\subsection{Ancilla Based Quantum Circuits}
It is well-known that using an ancilla register  eases the  complexity in the implementation of quantum gates (e.g. see \cite{Lanyon,Bary}). A particular type of ancilla based circuits can be described for the matrices represented as a sum of unitary matrices \cite{Daskin2012universal, Berry2015Taylor}:
Consider a unitary matrix, $A$, which is of dimension $N=2^n$ and represented as a sum of unitary processes, $U_i$s:
\begin{equation}
A=\sum_{i=0}^{M-1} U_i.
\end{equation}
A quantum operation can be described as a matrix vector transformation: For instance, let $\ket{in}$ represent a general input state. The application of $A$ on this input state can be described as $A\ket{in}$.  The result of this transformation can be obtained on an ancillary based circuit consisting of an ancilla and a system registers by using $U_i$s. 
One way of doing this with  an acilla register composed of $m=logM$ qubits can be described as follows: 
\begin{itemize}
\item First, all $U_i$s are aligned on the system register composed of $n$ qubits. 
\item Then, the application of each $U_i$ is controlled  by the ancilla register. 
Here,  the control scheme is determined from the binary representation of $i$ in a way that $U_i$ is applied when the ancilla is in \ket{\bf i} state. Here, \ket{\bf i} is the $i$th vector in the standard basis.
This circuit can be represented in matrix form by a block diagonal matrix, $V$:
\begin{equation}
\label{EqVgeneral}
V=\left(\begin{matrix}
U_0 & & \\
& \ddots& \\
  & & U_{M-1}
\end{matrix}\right).
\end{equation} 
\item The initial state of the ancilla register  is assumed to be \ket{\bf 0} state. Then, it is put into the superposition state by the application of  a series of Hadamard gates, $H^{\otimes m}$.  Thus, the whole circuit in matrix form becomes  equal to the following: 
\begin{equation}
\mathcal{U}=
(H^{\otimes m}\otimes I) V = \frac{1}{\sqrt{M}}\left(\begin{matrix}
\sum_{i=0}^{M-1} U_i & J_1\\
J_2& J_3
\end{matrix}\right),
\end{equation}
where $I$ is the identity matrix and $J_1, J_2$, and $J_3$ represent the other parts of $\mathcal{U}$.  
\end{itemize} 

$\mathcal{U}$ can be employed to emulate the action of $A$ with the success probability $\frac{1}{M}$:  If we apply $\mathcal{U}$ to the input \ket{\bf 0}\ket{in}; in the output, for the states in which the ancilla is equal to \ket{\bf 0}, the circuit gives the action of $A$ on the system register, $A\ket{in}$. 
 Since $A$ in $\mathcal{U}$ is normalized by the coefficient $\frac{1}{\sqrt{M}}$, the probability to see the ancilla in \ket{\bf 0} is $\frac{1}{M}$. 
 
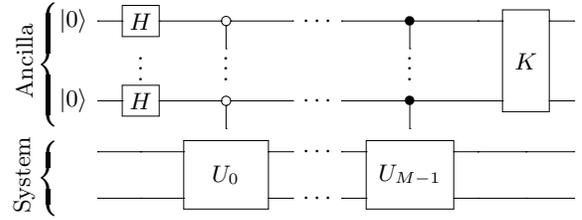
\begin{figure}
\[ \Qcircuit @C=1em @R=0.5em {
	&		&		&	&		&		&		&		&	&		&		&			&		&	\\
	&		&	\ket{0}	&	&	\gate{H}	&	\ctrlo{1}	&	\qw	&	\cdots	&	&	\ctrl{1}	&	\qw	&	 \multigate{4}{K} 		&	\qw	&	\\
	&		&		&	&		&		&		&		&	&		&		&			&		&	\\
\rotatebox[origin=c]{90}{Ancilla}	&	\rotatebox{270}{$\underbrace{\hspace{5em}}$}	&		&	&	\vdots	&	\vdots	&		&		&	&	\vdots	&		&			&		&	\\
	&		&		&	&		&		&		&		&	&		&		&			&		&	\\
	&		&	\ket{0}	&	&	\gate{H}	&	\ctrlo{1}	&	\qw	&	\cdots	&	&	\ctrl{1}	&	\qw	&	\ghost{K}		&	\qw	&	\\
	&		&		&	&		&		&		&		&	&		&		&			&		&	\\
	&		&		&	&	\qw	&	 \multigate{2}{\ \ U_0\ \ } 	&	\qw	&	\cdots	&	&	 \multigate{2}{U_{M-1}} 	&	\qw	&	\qw		&	\qw	&	\\
\rotatebox{90}{System}	&	\rotatebox{270}{$\underbrace{\hspace{3em}}$}	&		&	&		&		&		&		&	&		&		&			&		&	\\
	&		&		&	&	\qw	&	 \ghost{\ \ U_0\ \ }	&	\qw	&	\cdots	&	&	 \ghost{U_{M-1}}	&	\qw	&	\qw		&	\qw	&	\\
}\]
\caption{The circuit representation of $\mathcal{U}$.}
\label{FigCircuit1General}
\end{figure}

This can be also generalized to $A=\sum_{i=0}^{M-1} k_i U_i$ by using an operator $K$ whose first row and column are equal to the coefficients $k_0, \dots, k_{M-1}$ (Note that the coefficients are assumed to be normalized.):
\begin{equation}
\label{EqGeneralU}
\mathcal{U}=
(H^{\otimes m}\otimes I) V (K\otimes I)= \frac{1}{\sqrt{M}}\left(\begin{matrix}
\sum_{i=0}^{M-1} k_iU_i & J_1\\
J_2& J_3
\end{matrix}\right).
\end{equation}
$K$ can be considered as a Householder transformation  for the vector $\bra{\bf k}=\left(k_0, \dots, k_{M-1} \right)$: i.e., $K\ket{\bf k}=\ket{\bf 0}$ and $K\ket{\bf 0}=\ket{\bf k}$. The circuit representing $\mathcal{U}$ is drawn in Fig.\ref{FigCircuit1General}.

\subsection{Amplitude Amplification}
As derived from Eq.\eqref{EqGeneralU},  the success probability of  emulating the unitary matrix $A$ by the circuit, $\mathcal{U}$, is   $\frac{1}{M}$. The probability depends on the size of the ancilla and decreases exponentially in the number of qubits contained in the ancilla. 
This can be increased upto one by the application of the  amplitude amplification, which is the main part of the 
famous Grover's search algorithm \cite{Grover1998} finding a solution encoded in a quantum state.

The amplitude amplification is described as follows  \cite{Mosca1998quantum, Brassard2002,Kaye2006}:
Let the state \ket{\psi} be defined as the linear combination of the standard basis resulted from the application of the operator $B$ to an initial zero state:
\begin{equation}
\label{EqPsiinStandard}
\ket{\psi}=  B\ket{\bf 0}=\sum_{x=0}^{N-1}b_x\ket{\bf x},
\end{equation}
where $b_x$ is the amplitude of the $x$th vector, \ket{\bf x},  in the standard basis. Let also some states in \ket{\psi} be considered as the ``good" states represented by the set $X_{good}$, and the remaining states be the ``bad" states represented by $X_{bad}$:
\begin{equation}
\label{EqPsiinGoodBad}
\ket{\psi}=  \sum_{x\in X_{good}}b_x\ket{\bf x} + \sum_{x\in X_{bad}}b_x\ket{\bf x}.
\end{equation}
The re-normalizations of  $X_{good}$ and $X_{bad}$ form a new two dimensional basis set \{\ket{\psi_{good}},\ket{\psi_{bad}}\}. \ket{\psi} can be rewritten in this new basis with coefficients represented by $sin(\theta)$ and $cos(\theta)$ with $\theta\in[0,\pi/2]$:
 \begin{equation}
\ket{\psi}= sin(\theta)\ket{\psi_{good}}+cos(\theta)\ket{\psi_{bad}}.
\end{equation}
Here, $sin(\theta)^2=P_{good}=\sum_{x\in X_{good}}|b_x|^2 $, and $cos(\theta)^2=P_{bad}=\sum_{x\in X_{bad}}|b_x|^2$.

In the amplitude amplification algorithm, the amplitude of \ket{\psi_{good}} is amplified by using the following search iterate:
\begin{equation}
\label{EqUpsiUf}
Q=U_{\psi}^{\perp} U_f.
\end{equation} 
Here, when applied to \ket{\psi}, $U_f$  negates the sign of \ket{\psi_{good}} and does nothing to \ket{\psi_{bad}}:
 \begin{equation}
U_f\ket{\psi}=-sin(\theta)\ket{\psi_{good}}+cos(\theta)\ket{\psi_{bad}}.
\end{equation}
 When  $\ket{\psi_{good}}$ is known,  $U_f$ can be easily constructed in matrix form as follows:
\begin{equation}
\label{EqUfGeneral}
U_f=I-2\ket{\psi_{good}}\bra{\psi_{good}}.
\end{equation}

The other operator in the search iterate is $U_{\psi}^{\perp}$   described in matrix form as: 
\begin{equation} 
 U_{\psi}^{\perp}=2\ket{\psi}\bra{\psi}-I=BU_0^\perp B^{-1}, 
 \end{equation} 
where $U_0^\perp=(2\ket{\bf 0}\bra{\bf 0}-I)$ is the reflection about the axis orthogonal to \ket{\bf 0}, and $B^{-1}$ is the inverse of the matrix $B$. When $B$ is unitary, $B^{-1}$ is equal to the conjugate transpose of $B$, $B^\dagger$.

It is easy to prove that 
 when $Q$ is applied $k$ number of times to \ket{\psi}, the final state ends up in the following:
\begin{equation}
Q^k\ket{\psi} =
 sin\left((2k+1)\theta\right)\ket{\psi_{good}}
+\cos\left((2k+1)\theta\right)\ket{\psi_{bad}}.
\end{equation}
From the above equation,  the highest probability for the good states, $ sin\left((2k+1)\theta\right)^2$, can be obtained when $(2k+1)\theta \approx\frac{\pi}{2}$. This leads $O(\frac{\pi}{4\theta})$ as a bound for $k$.

\subsubsection{Application to $\mathcal{U}$ and Oblivious Amplitude Amplification}
In the circuit $\mathcal{U}$ given in Fig.\ref{FigCircuit1General}, the good and the bad states are already known: the set of good states are those where the ancilla qubit is in \ket{\bf 0} state and the rest is the set of the bad states. Therefore, the iteration operator used in the amplitude amplification can be determined.

Let \ket{\beta} be the output state of $\mathcal{U}$.
\ket{\beta} can be written in terms of the combination of the good and the bad states:
\begin{equation}
\label{EqBetaGoodBad}
\ket{\beta}=\frac{1}{\sqrt{M}}\ket{\beta_{good}}+\frac{\sqrt{M-1}}{\sqrt{M}}\ket{\beta_{bad}}.
\end{equation}
Since the good states are the ones where the ancilla register equal to zero, the normalized $\ket{\beta_{good}}$ reads the following:
\begin{equation}
\label{EqBetaGood}
\ket{\beta_{good}}=\ket{\bf 0}\sum_{x=0}^{N-1}\beta_x\ket{\bf x},
\end{equation}
where \ket{\bf x} is the $x$th vector in the standard basis. 

It is shown that the amplitude amplification can be done on the first register by using the following search iterate \cite{Paetznick2014repeat, Berry2014exponential,Berry2015Taylor}. 
\begin{equation}
\label{EqQ1}
 Q= -\mathcal{U}S\mathcal{U} S,
\end{equation} 
where $S$ is considered in place of $U_f$ and has the same effect as $U_{0}^\perp$ in the iteration: 
\begin{equation}
\label{EqSExplicit}
 S= \left(I^{\otimes^m}-2 \ket{\bf 0}\bra{\bf 0}\right) \otimes I^{\otimes^n}.
 \end{equation}
 The main difference of this iteration from the standard amplitude amplification is the independence from the input to the system register: $\mathcal{U}$  does not include the input preparation operator which converts a zero state into the desired input. 

Since the amplitude of the good states is $1/\sqrt{M}$; $Q$ is applied exactly $\left \lfloor{\frac{\pi}{4}\sqrt{M}} \right \rfloor$ number of times to maximize the amplitude of  $\ket{\beta_{good}}$.  

\section{Application in the Case $A$ is  Non-Unitary}
When $A$ is non-unitary; the expected output from the circuit $\mathcal{U}$ can no longer be described by the fixed amplitudes given in Eq.\eqref{EqBetaGoodBad} since it changes based on the input state to the system. This necessitates the inclusion of the input preparation  in the operator $U_{\psi}^{\perp}$ of the standard amplitude amplification:
\begin{equation}
 U_{\psi}^{\perp}=\mathcal{U} (I^{\otimes m} \otimes A_{in}) U_0^\perp (I^{\otimes m} \otimes A_{in})^{\dagger}\mathcal{U}^{\dagger},
\end{equation}
where $A_{in}$ converts the zero state into the desired input state. 
To reach the maximum amplitude,  in this case, $Q$ needs to be applied $\left \lfloor{\frac{\pi}{4\eta}\sqrt{M}} \right \rfloor$ number of times, where $\eta$ is the norm of the expected output state produced from $||UU_{in} \ket{\bf 0}||$. Since $\eta$ is not known, one can apply quantum search algorithm \cite{Kaye2006}. In that case, the first register is partially measured: If it is zero, then the algorithm is stopped. Otherwise, the iteration operator $Q$ is applied again.

In the following subsections, we shall show how to extend $A$ to a  unitary matrix in order to avoid input dependence in the amplitude amplification and successfully apply the oblivious amplitude amplification.

\subsection{Extending $A$ to a Unitary Matrix}
It is known that every Hermitian positive definite matrix has a unique Hermitian positive definite square root which can be numerically  computed via Schur decomposition \cite{Bjorck1983schur, Higham1987computing}. 
Using the square root of $(I-A^2)$, a non-unitary Hermitian matrix, $A$, can be extended to a unitary, $U$,  as in the following form:
\begin{equation}
\label{EqUwithsqrt}
U =\left(\begin{matrix}
A&\sqrt{I-A^2}\\
\sqrt{I-A^2} &-A
\end{matrix}\right).
\end{equation}
It is not difficult to see that $U$ is a unitary matrix: $UU^\dagger = U^\dagger U=I^{\otimes n+1}$. This can be seen easier when $U$  is written in the cosine-sine decomposition as:
\begin{equation}
U =\left(\begin{matrix}
A&\sqrt{I-A^2}\\
\sqrt{I-A^2} &-A
\end{matrix}\right)=
\left(\begin{matrix}
R&0\\
0 &R
\end{matrix}\right)
\left(\begin{matrix}
C&S\\
S&-C
\end{matrix}\right)
\left(\begin{matrix}
R^\dagger&0\\
0 &R^\dagger
\end{matrix}\right),
\end{equation}
where  the eigendecomposition of $A$ is described by $A=RCR^\dagger$, and $C$ and $S$ are the diagonal matrices with the diagonal elements cosine and sine of the eigenvalues of the matrix $A$. 
{Note that this cosine-sine decomposition can also be used to generate a quantum circuit for $U$ in terms of single and two qubit gates \cite{, Tucci1999rudimentary,Mottonen,Shende2006}. }

\subsubsection{Simulation of $U$}
Computation of $\sqrt{I-A^2}$  requires the eigendecomposition  of $I-A^2$. 
Since the eigenvalues of $\sqrt{I-A^2}$ and $A$ are related as cosine and sine of some values, the eigendecomposition of $A$ is enough to form the matrix $\sqrt{I-A^2}$. 
Eigendecomposition of a matrix can be found through numerical methods such as power iterations and QR method. 
For a Hermitian matrix whose size is reasonable for classical computers,  obtaining a kind of approximate decomposition is feasible in general. 
However, when the system size is huge, this problem becomes a very heavy computational task for the classical computers so that, in some cases, decomposition may be impossible (note that the computational complexity in terms of the system size is generally $O(N^3)$.)\cite{Parlett1980symmetric}.

Therefore, when the computation of the eigendecomposition of $A$ is available;
  one can use the matrix $A$  on quantum computers as follows: 
  In the circuit for $U$, the initial state of the first qubit is always set to \ket{0}. 
  Thus, we apply $U$ to an initial state \ket{input}=\ket{0}\ket{in}, where \ket{in} represents the initial input desired to be applied to $A$.  This generates the following output vector:
\begin{equation}
\ket{output}=\left(\begin{matrix}
A&\ket{in}\\
\sqrt{I-A^2}&\ket{in}
\end{matrix}\right)
\end{equation}  
To the above vector,  if we apply the operator $P=(\ket{0}\bra{0}\otimes I^{n})$, then the system collapses to the state $A\ket{in}$. 
Therefore, at the end of circuit $U\ket{input}$, the probability to see the first qubit in zero state, determines the success probability of the simulation done by the following process. 

When the eigendecomposition of $A$ is not available (because of the computational complexity, this may be the case when the system size of $A$ is large),  we shall try to estimate $U$ and process the estimated non-unitary on the ancilla based circuit described in the previous section.

\subsubsection{Estimation of $U$}
{From now on,  we assume that $A$ is a real symmetric matrix.} Then, we normalize the matrix $A$ as $A=\frac{A}{\mu}$ with $\mu=\sqrt{max_i\sum_j A_{ij}^2}$. This makes 2-norms of the columns of the matrix less than or equal to 1.
Then, instead of  the matrix in Eq.\eqref{EqUwithsqrt}, we construct the following matrix:
 \begin{equation}
 \label{EqConstructedNonUnitary}
U=\left(\begin{matrix}
A&D\\
D&-A
\end{matrix}\right),
\end{equation}
where $D$ is a diagonal matrix described by the diagonal entries $D_{ii}=\sqrt{1-\sum_j A_{ij}^2}$.  
The matrix $U$ is not a unitary matrix; however,  $A^2 + D^2$  and so $UU^\dagger=U^\dagger U = U^2$ have diagonal elements equal to one. { In addition, since $A$ is normalized with $\mu$; $\rho(A^2)\leq \rho(A)\leq 1$, 
where $\rho(A)$ is the spectral radius (the largest eigenvalue) of $A$. In particular cases such as $\rho(A) < 1$ or a diagonally dominant $A$, the off-diagonal entries of $A^2$ are likely to be small, and so one can expect the matrix $U$ to be close to a unitary since $A^2 + D^2$ becomes close to an identity matrix.}

We shall describe the closeness of $U$ to a unitary matrix via polar decomposition which is known to produce the closest unitary matrix to a given arbitrary matrix (e.g. see  \cite{Higham1988matrix}):
In the polar decomposition, $U=\widetilde{U}\widetilde{H}$, while $\widetilde{U}$  is the closest unitary to $U$, $\widetilde{H}$ is an Hermitian positive definite matrix. 
 We define the following as a measure of the closeness of the estimated unitary $U$ to $\widetilde{U}$:
\begin{equation}
\label{Eqc}
c=\frac{||U-\widetilde{U}||^2}{||U||^2}=\frac{||\widetilde{H}-I||^2}{||\widetilde{H}||^2},
\end{equation}
{where $||.||$ represents the norm of a matrix. Also note that a similar measure, ${||U-\widetilde{U}||}$, is used by Berry et al. as an error bound in the modified version of the oblivious amplitude amplification \cite{Berry2015Taylor}.
 To bound the value of $c$, we will use the Frobenius norm:
$||.||_F$, which is also invariant under unitary transformations. Thus, $||\widetilde{H}||_F =  ||U||_F = \sqrt{trace(UU^\dagger)}=\sqrt{trace(\widetilde{H}^2)}=\sqrt{2N}$. 
In addition, we can represent the term $||\widetilde{H}-I||_F$ as follows:
\begin{equation}
\begin{split}
||\widetilde{H}-I||_F = & \sqrt{trace(\widetilde{H}^2-2\widetilde{H} + I)} \\
= &
\sqrt{trace(\widetilde{H}^2)-trace(2\widetilde{H}) + trace(I))}
\\
= & \sqrt{4N-trace(2\widetilde{H})}
= \sqrt{4N - \Phi}.
\end{split}
\end{equation}
From the generalized mean equality; since the trace is the sum of eigenvalues, the following inequality holds:
 $\frac{1}{2N}trace(\widetilde{H}) 
 \leq \sqrt{\frac{1}{2N}trace(\widetilde{H}^2)}$. 
As a consequence of this, 
 $\Phi=trace(2\widetilde{H}) \leq 4N$ and so  $\sqrt{4N - \Phi} \leq 2N$.
This makes $c$ in terms of the Frobenius norms  always less than or equal to one. 
When $\Phi=4N$,  $c$ becomes zero and so the matrix can be considered a unitary matrix. 
Note that one can also try to maximize $c$ for different choices of $A$.}

{In the numerical examples, we shall compute $c$ by using 2-norms of the matrices instead of the Frobenious norms: Since the error also depends on the input state,  the 2-norm obtained from the maximum eigenvalue likely gives a better approximation for the possible maximum error.}
Furthermore, we shall use the following as an estimate lower bound for the fidelity of the output after the oblivious amplitude amplification algorithm:
\begin{equation}
\label{EqEstimatedFidelity}
ef =|1-c|^2.
\end{equation}
The estimated fidelity would be one for the unitary matrices. And it would be small when the closeness between $U$ and $\widetilde{U}$ is small.


When $U$ is non-unitary, it cannot be used directly on quantum computers. 
Berry et al.\citep{Berry2015Taylor} have proved that a slightly modified version of the oblivious amplitude amplification works also for any arbitrary non-unitary matrix which is close to a unitary matrix. Therefore, since the constructed matrix in Eq.\ref{EqConstructedNonUnitary} is expected to be close to a unitary matrix, the modified oblivious amplitude amplification can be used for the simulation of this matrix in the ancilla based circuit. However, in the next section, the numerical results for the randomly generated matrices show that one can use the oblivious amplitude amplification given in Eq.\eqref{EqQ1} without modification  for the  matrices generated by \eqref{EqConstructedNonUnitary}.
{Note that  the modified version of the oblivious amplitude amplification would give better results; however, it requires a projection operator.
}
\subsection{Numerical Examples}
In the numerical examples, we shall use the standard oblivious amplitude amplification given in Eq.\eqref{EqQ1} along with the circuit design method described in Ref.\cite{Daskin2012universal}. The circuit design method is only included for the completeness. One can also run simulations by generating random matrices in the form Eq.\eqref{EqConstructedNonUnitary} and then writing each one of them as a direct sum of $M$ unitary matrices.

\subsubsection{General Circuit Designs}
 An ancilla-based quantum circuit design technique  is introduced in Ref. \cite{Daskin2012universal}.  A $2n$-qubit system  $\mathcal{U}$ is employed to emulate a general $n$-qubit quantum operation, $U$. 
The primary
intuition of the technique is based on generating the matrix-vector product result of a quantum operation inside the new system. This intuition is conveyed in two parts: First,  the ancilla register is put into the superposition state, then  the following matrix $V$ similar to the one in Eq.\eqref{EqVgeneral} is used:
\begin{equation}
\label{EqV}
V=\left(\begin{smallmatrix}
\begin{smallmatrix}
\alert{u_{11}}&\alert{u_{12}}&\alert{\cdots} &\alert{ u_{1N}}\\
\bullet&\bullet&\dots&\bullet\\
 \vdots&\vdots&\ddots&\vdots\\
 \bullet&\bullet&\dots&\bullet
\end{smallmatrix}& & &
\\
&
\begin{smallmatrix}
\alert{u_{21}}&\alert{u_{22}}&\alert{\dots }& \alert{u_{2N}}\\
\bullet&\bullet&\dots&\bullet\\
 \vdots&\vdots&\ddots&\vdots\\
 \bullet&\bullet&\dots&\bullet
\end{smallmatrix}&&
\\&
&  \ddots  &
\\
 & & &\begin{smallmatrix}
\alert{u_{N1}}&\alert{u_{N2}}&\alert{\dots} & \alert{u_{NN}}\\
\bullet&\bullet&\dots&\bullet\\
 \vdots&\vdots&\ddots&\vdots\\
 \bullet&\bullet&\dots&\bullet
\end{smallmatrix}
\end{smallmatrix}\right),
\end{equation}
where $u_{ij}$s are the matrix elements of $U$, and  each ``$\bullet$" represents a matrix element which plays no significant role in the description.
If this matrix is applied to a quantum state in which the ancilla is in the superposition state, the desired output is obtained in the states where the system register is in \ket{\bf 0} state:
\begin{equation}
\label{EqExtendedV}
V(\frac{1}{\sqrt{M}}\sum_{x=0}^{M-1}\ket{\bf x}\ket{input})
= \frac{1}{\sqrt{M}}\left(\begin{matrix}\beta_{0}\\ \vdots\\
\beta_i\\\vdots\\ \beta_{N-1}\\\vdots\end{matrix}\right).
\end{equation}
Obviously, when we swap the ancilla and the system register, we obtain the same output as in Eq.\eqref{EqBetaGoodBad} with the good states in Eq.\eqref{EqBetaGood} (Note that here $M=N$.). As a consequence, the same amplitude amplification given in Eq.\eqref{EqQ1} can be applied to increase the probability upto one.

Having a non-unitary $U$ does not change the circuit design. However, it changes the success probability given in the above equation. When $U$ is close to a unitary matrix, we can apply the oblivious amplitude amplification  in Eq.\eqref{EqQ1} (One can also use the modified version of the oblivious amplitude amplification given in Ref.\cite{Berry2015Taylor}).

\subsubsection{Numerical Examples}

In the numerical examples, the oblivious amplitude amplification given in Eq.\eqref{EqQ1} is applied to the non-unitary matrices constructed in the form given in Eq.\eqref{EqConstructedNonUnitary}. 
The following method is used to generate random symmetric matrices:
\begin{itemize}
\item An $N\times N$ symmetric random matrix, $A$, with the elements in $[-1, 1]$ is generated.
\item Then, $A$ is normalized by $\mu=\sqrt{max_i\sum_j A_{ij}^2}$.
\item Then, the diagonal matrix $D$ is obtained: $D_{ii}=\sqrt{1-\sum_j A_{ij}^2}$.
\item Then, the matrices A and D are used to construct $2N\times 2N$ matrix $U$. 
\end{itemize}

In addition, we also generate random input state to $U$ in the form: $\ket{input}=(\ket{0}\otimes \ket{in})$, where \ket{in} is a random quantum state with real coefficients.

We have run the simulations for 100 random cases with $U$  of orders $N=16$, $N=32$, $N=64$, and $N=128$ (Note that the total system sizes with the ancilla register becomes 
$(2^8\times2^8), (2^{10}\times2^{10}), (2^{12}\times2^{12})$ and $(2^{14}\times2^{14})$, respectively.).  For each random case, the estimated fidelity given in Eq.\eqref{EqEstimatedFidelity}  has been computed as follows:
\begin{itemize}
\item First, to find the polar decomposition of $U$, the singular value decomposition of $U$ is computed:
$U=V_1E V_2^\dagger$, where $E$ represents the singular values and $V_1$ and $V_2$ are the right and left singular vectors.
\item Then, the closest unitary is computed from $\widetilde{U}=V_1V_2^\dagger$.
\item Then, $c$ given in Eq.\eqref{Eqc} is computed by using the 2-norms of the matrices.
\item Finally, the estimated fidelity is computed from Eq.\eqref{EqEstimatedFidelity}.
\end{itemize}
In the amplitude amplification process, the amplitudes of the first $N$ states-i.e., the states where the first register is in \ket{\bf 0}-are used to compute the success probability. Furthermore, the estimated fidelity is compared with the real fidelity which is the inner product of the collapsed quantum state (the first $N$ amplitudes) and the vector $(A\ket{in})$. 

\begin{figure*}
\subfloat[16x16 Matrix]{\includegraphics[width= 3.5in]{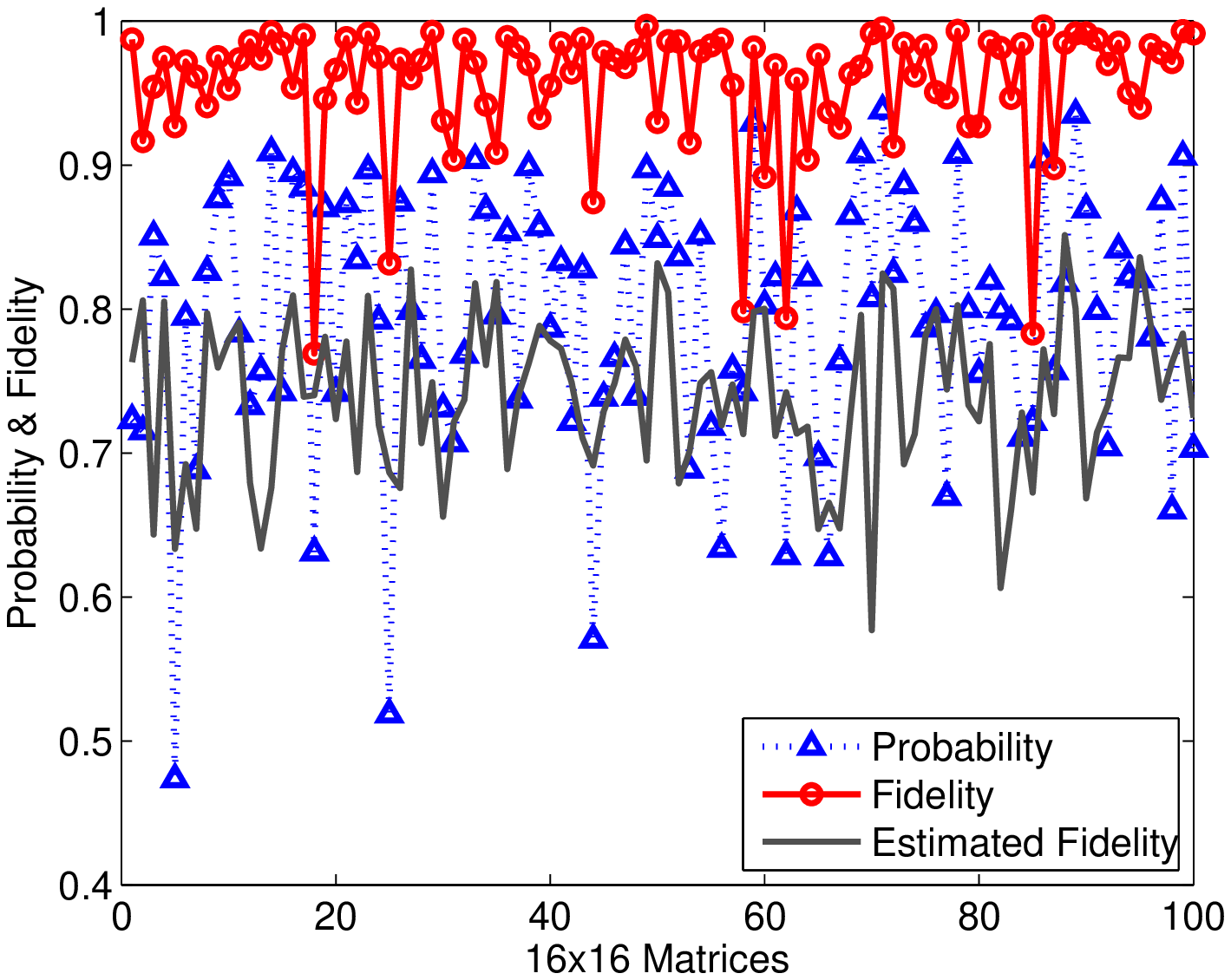}}
\subfloat[32x32 Matrix]{\includegraphics[width= 3.5in]{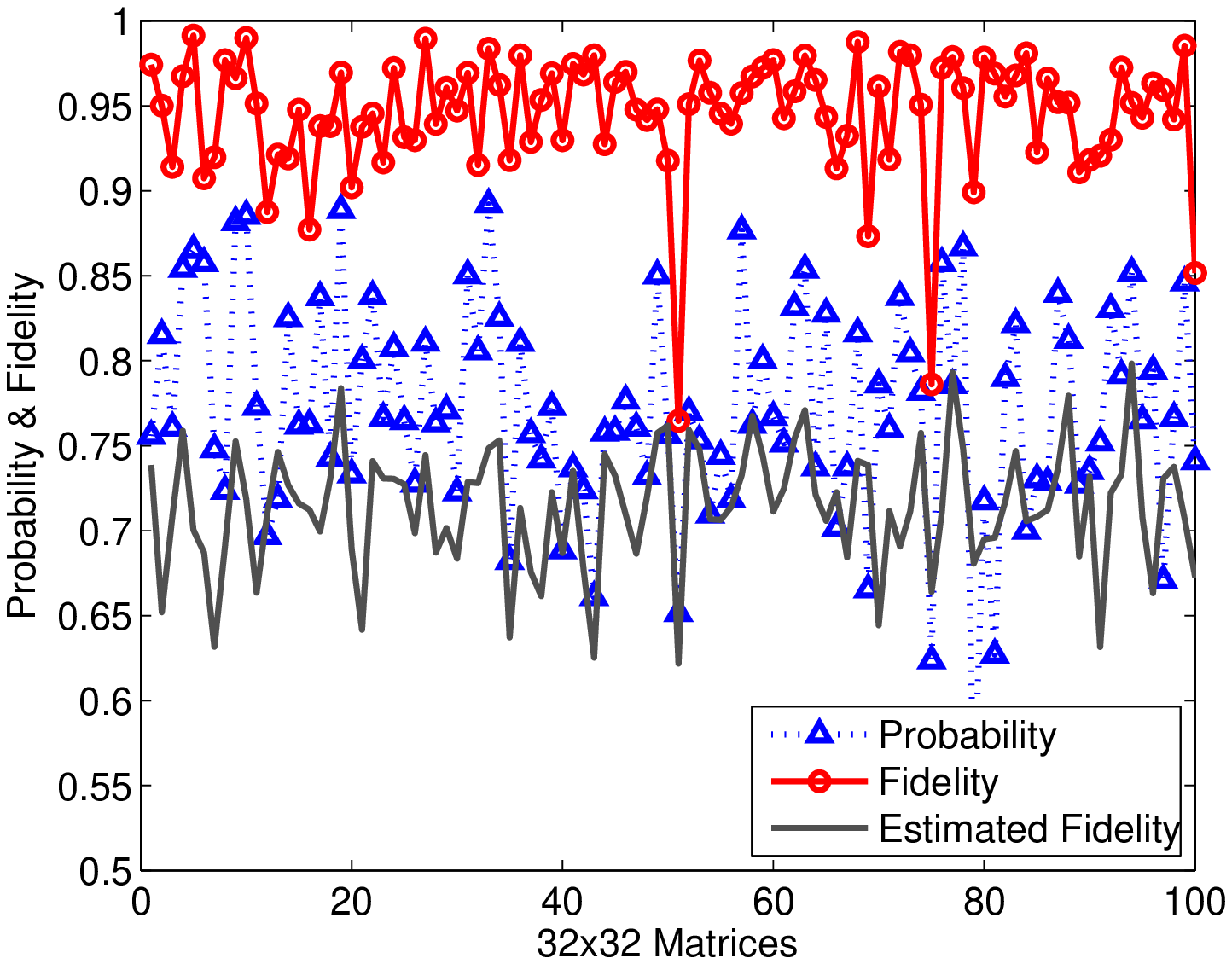}}\\
\subfloat[64x64 Matrix]{\includegraphics[width= 3.5in]{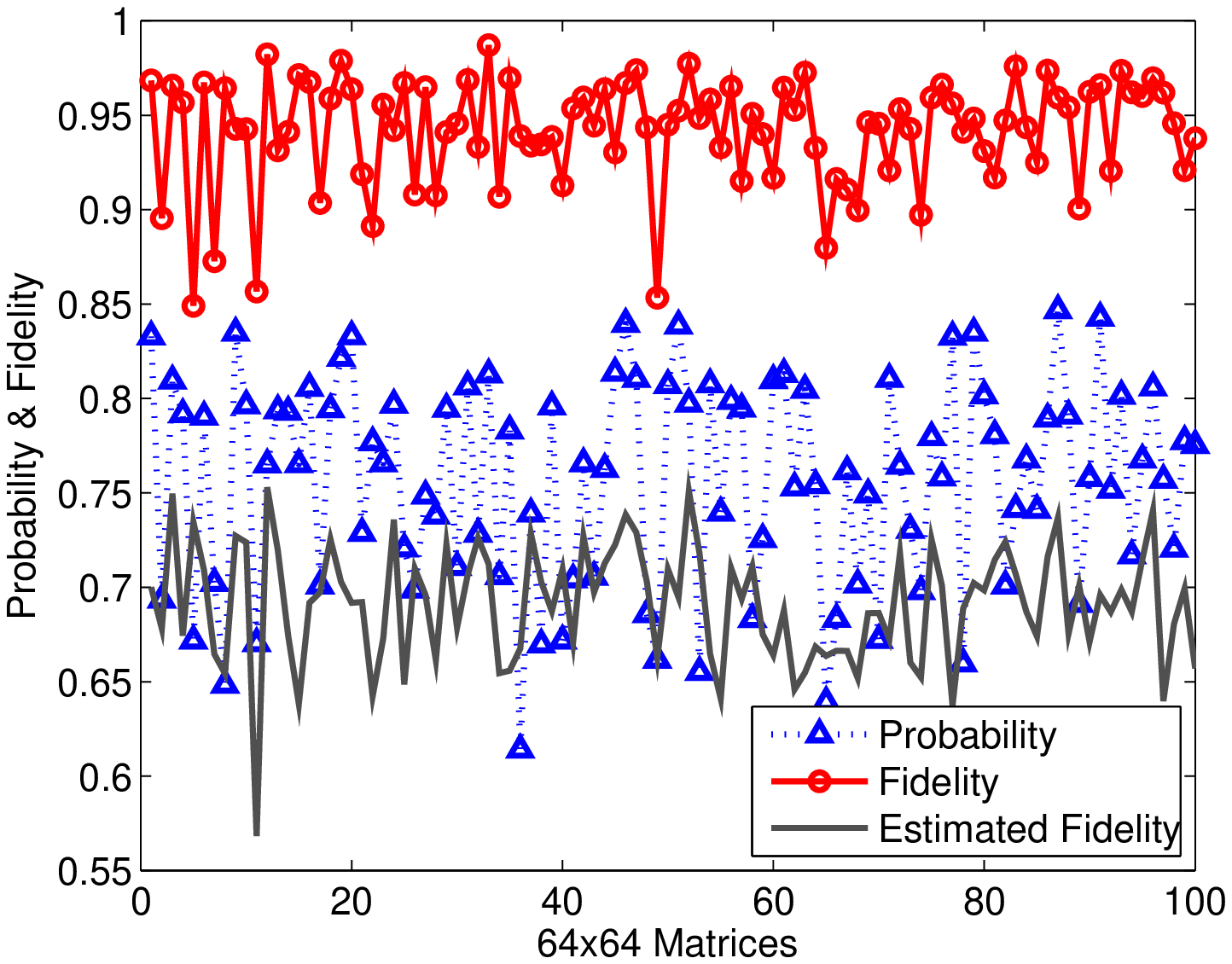}}
\subfloat[64x64 Matrix]{\includegraphics[width= 3.5in]{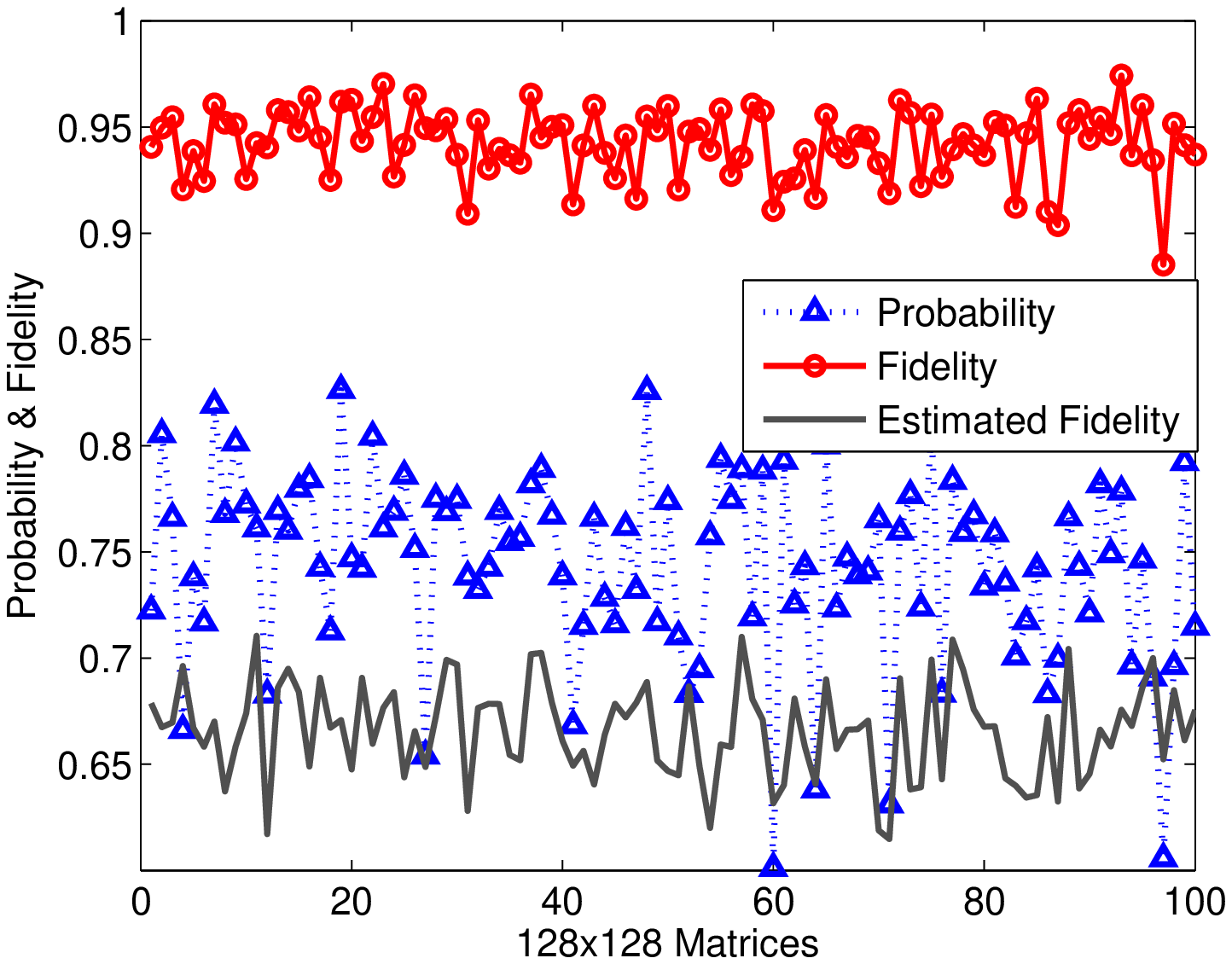}}
\caption{The change of the fidelity, the probability, and the estimated lowest fidelity  at the end of the amplitude amplification process for  100 random matrices of different sizes and random inputs. 
\label{figMatrices}
}
\end{figure*}

In Fig.\ref{figMatrices}, both $U$ and \ket{in} are generated randomly. And from the last iteration of the oblivious amplitude amplification (the state in which the success probability is the highest), the fidelity and probability are computed for each case. As can be observed in the figure, the mean value of the observed  fidelities is 0.9462 and the mean value of the observed success probabilities is 0.7666.

\begin{figure*}
\subfloat[16x16 Matrix]{\includegraphics[width= 3.5in]{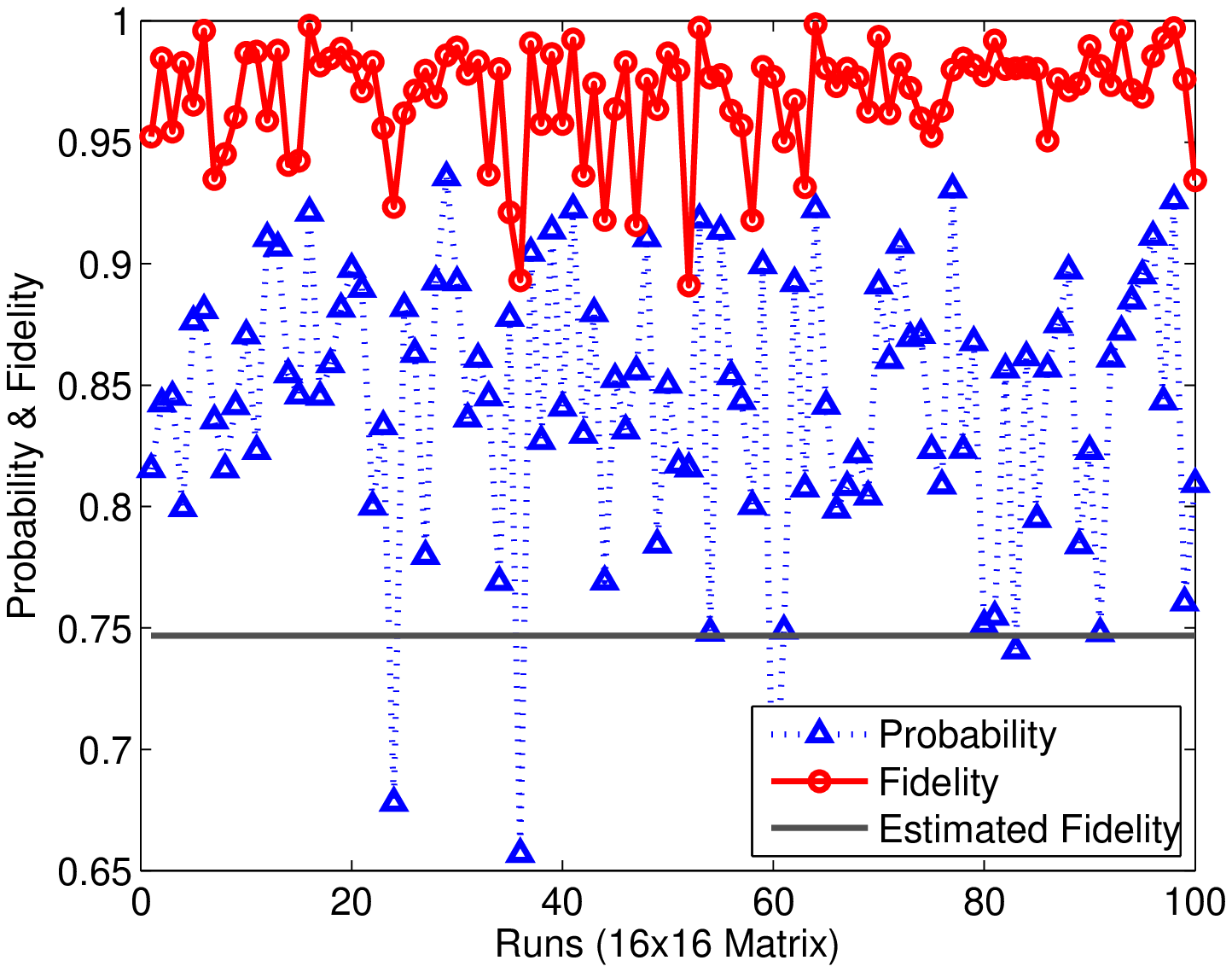}}
\subfloat[32x32 Matrix]{\includegraphics[width= 3.5in]{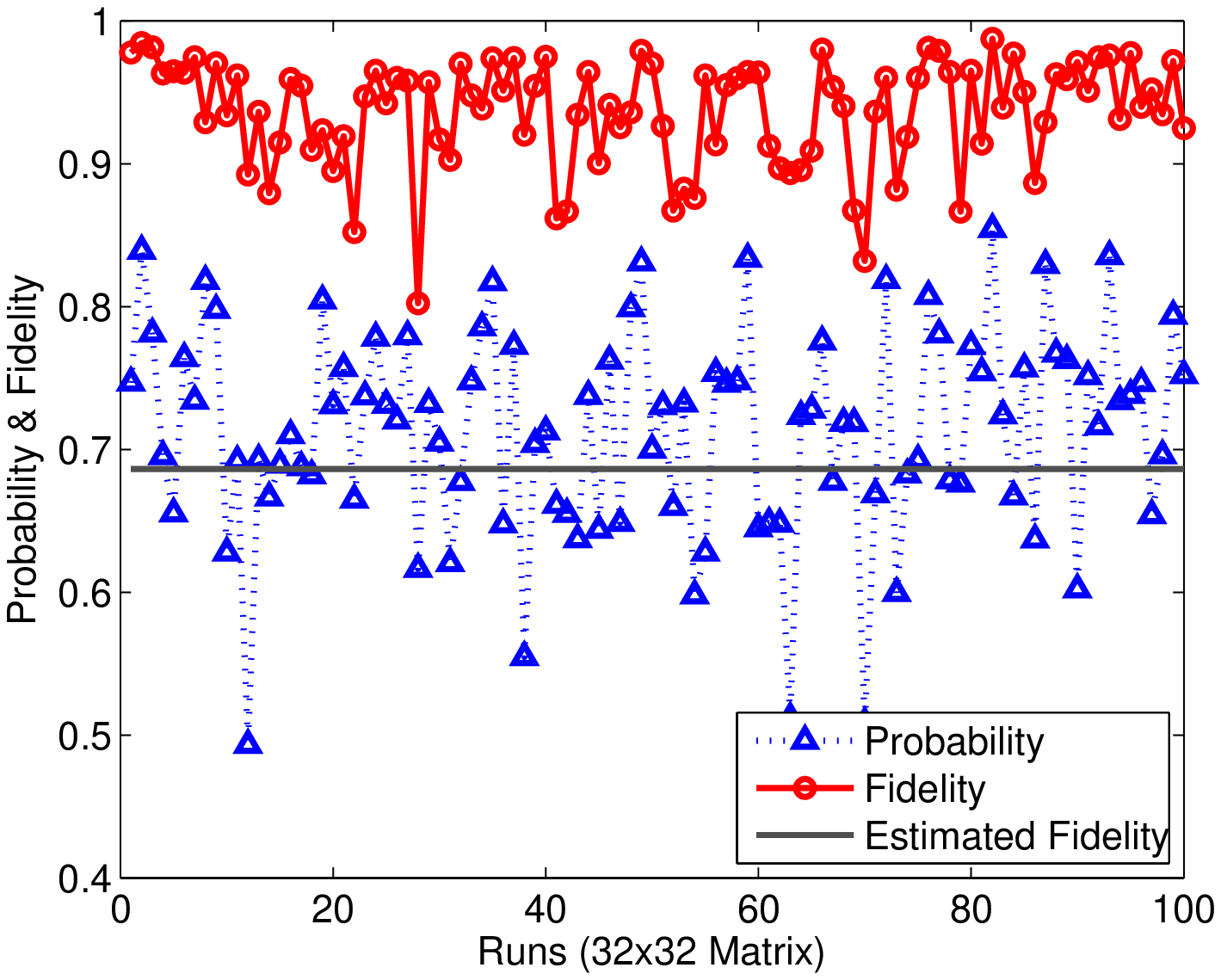}}\\
\subfloat[64x64 Matrix]{\includegraphics[width= 3.5in]{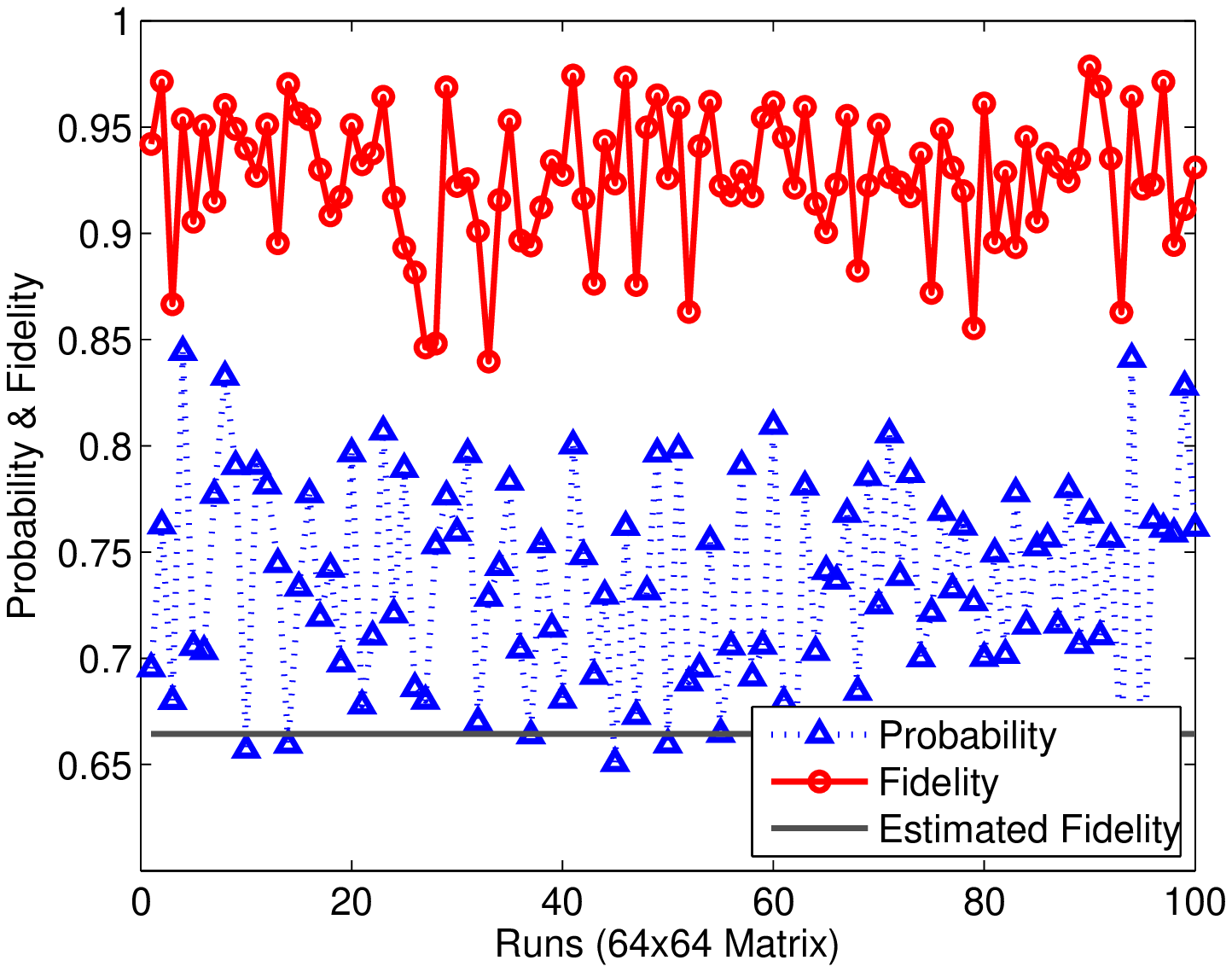}}
\subfloat[128x128 Matrix]{\includegraphics[width= 3.5in]{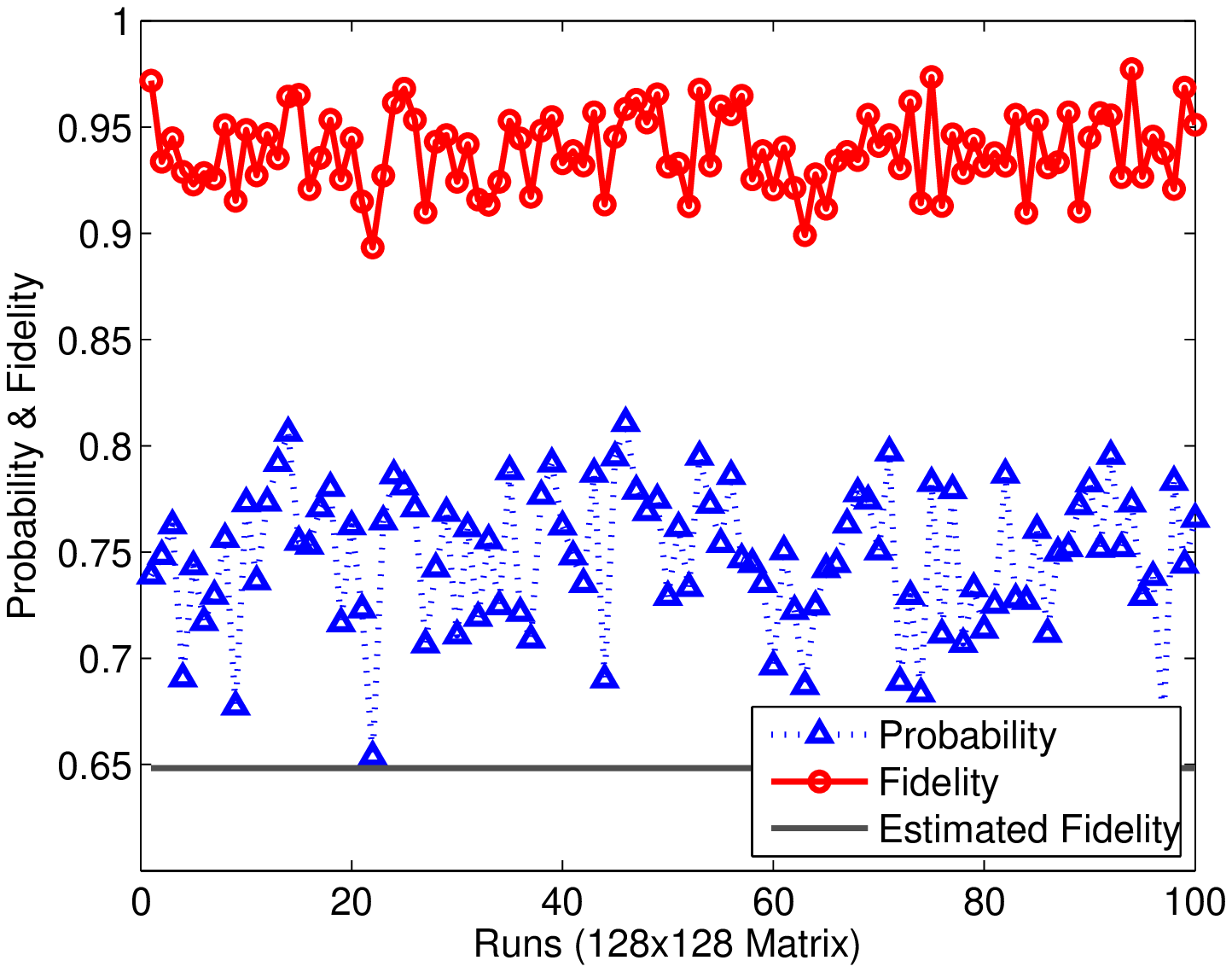}}
\caption{The change of the fidelity, the probability, and the estimated lowest fidelity  at the end of the amplitude amplification process. In each run, the matrix $U$ is unchanged but a different random input is used. 
\label{figRuns}
}
\end{figure*}

Furthermore, in Fig.\ref{figRuns}, the same matrix $U$ is tested in the amplitude amplification with  different input states in order to observe the effect of the change in the input state on the fidelity and  the validity of the estimated fidelity. 
Fig.\ref{figRuns} shows that the fidelity remains high and is always above the estimated fidelity (the mean of fidelities in all subfigures is 0.9426). Therefore,  the proposed estimated fidelity may provide a good lower bound for the fidelity. As in  Fig.\ref{figMatrices}, the mean value of the success probabilities here also remain as 0.7601.  


\begin{figure*}
\subfloat[16x16 Matrix]{\includegraphics[width= 3.5in]{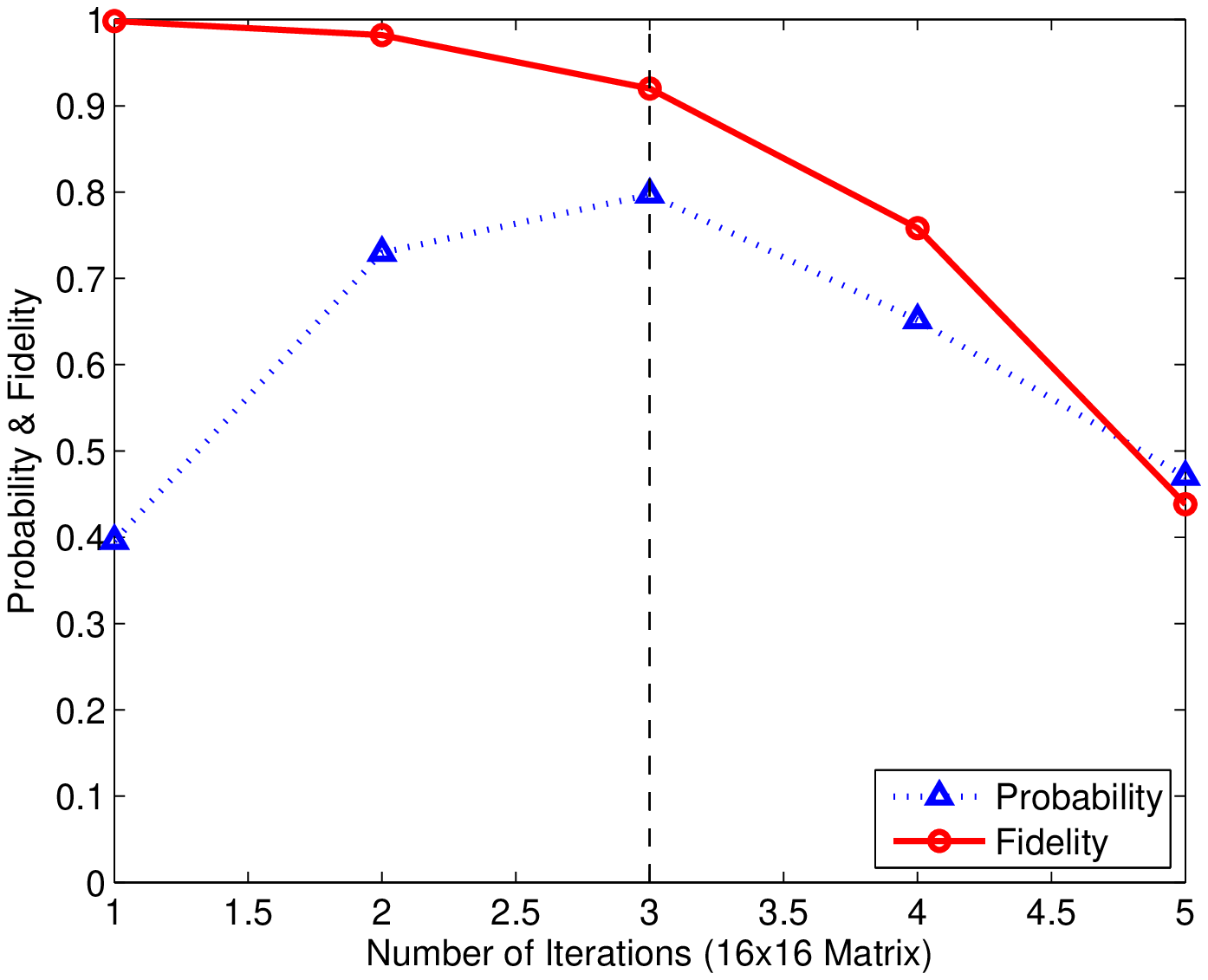}}
\subfloat[32x32 Matrix]{\includegraphics[width= 3.5in]{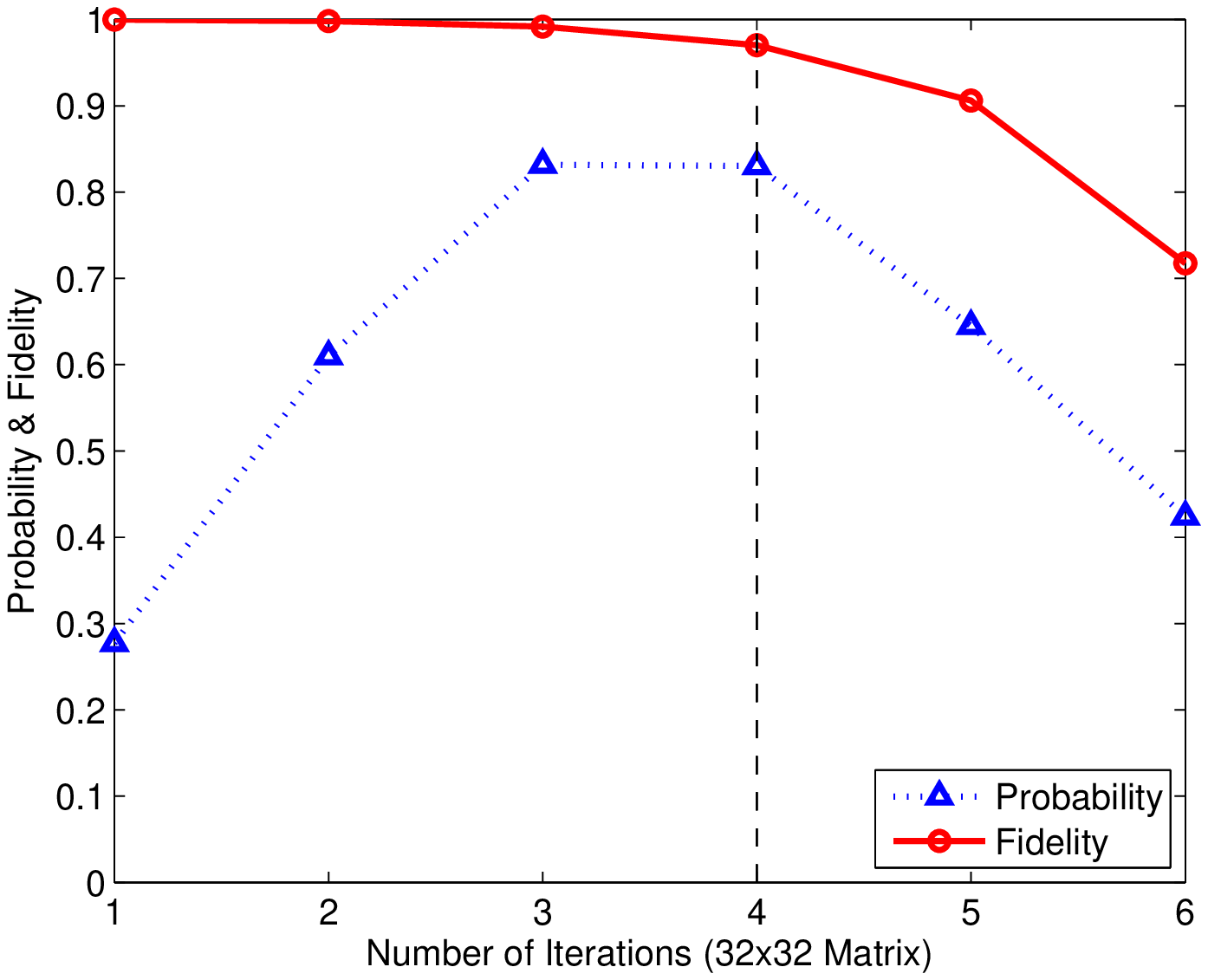}}\\
\subfloat[64x64 Matrix]{\includegraphics[width= 3.5in]{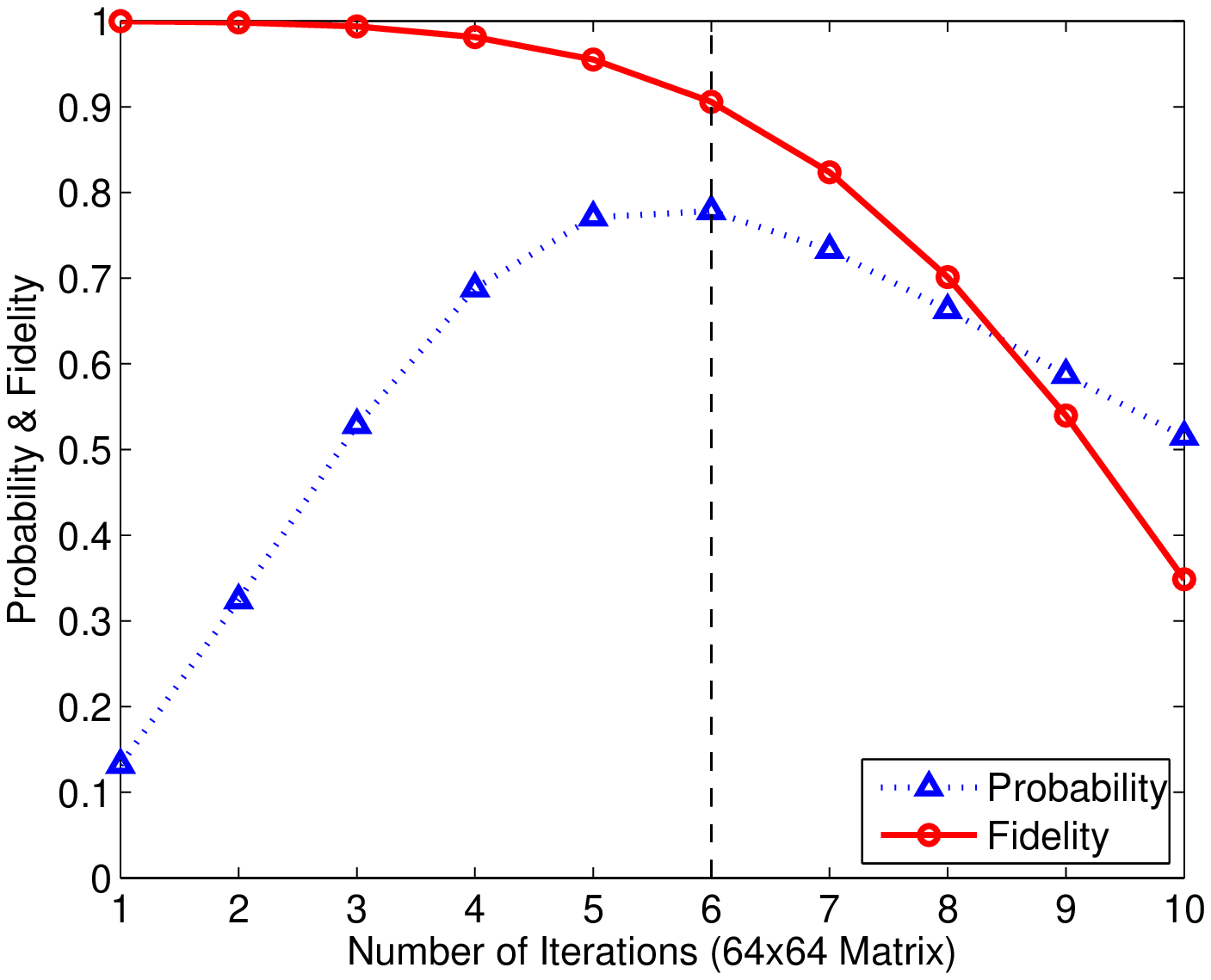}}
\subfloat[128x128 Matrix]{\includegraphics[width= 3.5in]{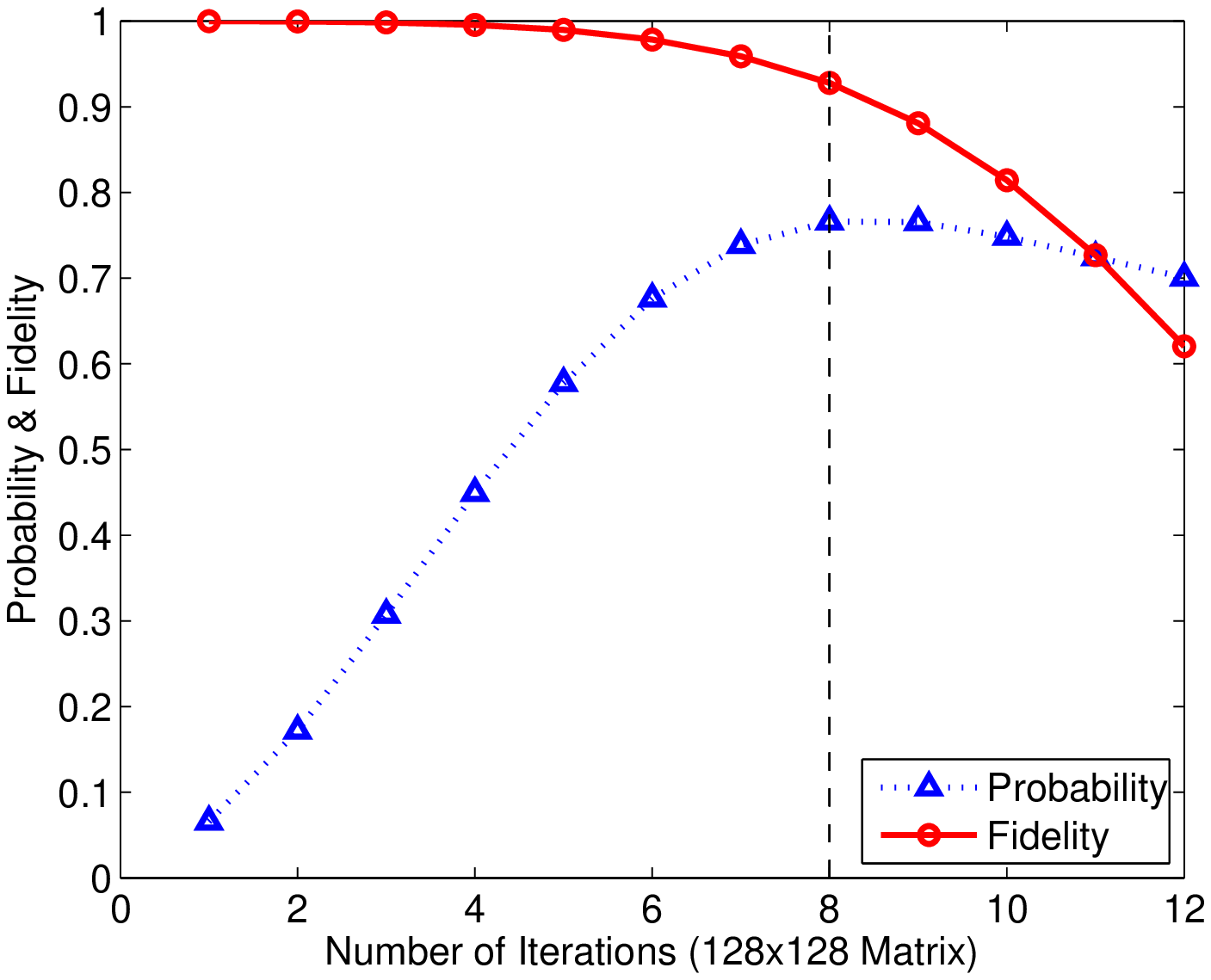}}
\caption{The change of the fidelity and the probability in the amplitude amplification process for  the random matrix $U$ given in Eq.\ref{EqConstructedNonUnitary} of different sizes. In each figure, the vertical dashed line shows the value of $k=\left \lfloor{\frac{\pi}{4}\sqrt{N}} \right \rfloor$.
\label{figIterations}
}
\end{figure*}

In Fig.\ref{figIterations}, we also present the change of the fidelity and the probability 
in the iterations of the oblivious amplitude amplification for random cases with the matrices of the same orders:  $N=16$, $N=32$, $N=64$, and $N=128$.  The required number of iterations for each case has been computed from $k=\left \lfloor{\frac{\pi}{4}\sqrt{N}} \right \rfloor$. For $N=16$, $N=32$, $N=64$, and $N=128$; respectively, $k=3$, $k=4$, $k=6$, and $k=8$ drawn  as dashed vertical lines on the figures. As seen in the figure: the fidelities in all figures remain high for a few iterations, but then start gradually decreasing while the probabilities are rising. When the probabilities reach their maximums either at the $k$th iteration or before that; the fidelities go down to around 0.95.  

As a general observation, we can see that  the numerical results presented in   Fig.\ref{figMatrices}, Fig.\ref{figRuns}, and Fig.\ref{figIterations} remain mostly similar despite  the change in the system size.  This indicates the independence of the results from the system size. 

\section{Discussion}
\label{SecDiscus}
\subsubsection{The Total Circuit Complexity}

Quantum circuit design methods can be broadly categorized into stochastic and non-stochastic approaches \cite{Daskin2014Diss}. 
In terms of stochastic approaches, generally genetic and evolutionary algorithms are used to obtain a circuit which minimizes an optimization function measuring the closeness of the found circuit to the given operator (e.g. \cite{Daskin2011decomposition}). {
Note that instead of designing a quantum circuit implementing a specific matrix, 
one can also consider the quantum control approach using optimization to find a way to implement the operator directly, without gates( e.g. \cite{Stojanovic2012quantum}). 
This can also be done for non-unitary processes.} 
On the other 
hand, non-stochastic deterministic methods are generally based on matrix decomposition techniques: for instance, methods based on the Cartan decomposition \cite{Love2008}, the cosine-sine decomposition \cite{Shende2006}, the QR decomposition \cite{Cybenko2001reducing} are presented. It is shown that a circuit for a general unitary matrix  requires at least $(2^{2n-2}-3n/4-1/4)$ 
number of two-qubit quantum gates \cite{Shende2006}.

The combination of the circuits for each $U_i$ forms a binary coded network controlled by the ancilla register. These types of networks can be decomposed into single and two qubit quantum gates by using a simple decomposition technique described in Ref.\cite{Mottonen}. 
The decompositions of the networks for the circuit design in Ref. \cite{Daskin2012universal} yield a circuit with $O(N^2-N)$ number of CNOT gates if $M=N$ (For  the complete complexity analyses, please refer to Ref.\cite{Daskin2012universal}.).  Therefore, the quantum complexity of the circuit $\mathcal{U}$ is $O(N^2-N)$. The amplitude amplification requires $O(\sqrt{N})$ repetitions which makes the total complexity $O(\sqrt{M}N^2-\sqrt{M}N)$. 

If each $U_i$ can be efficiently simulatable: i.e. each requires only $O(poly(n))$ gates; the total circuit complexity for $\mathcal{U}$ can be bounded by $O(poly(n)M)$: 
Since there would be $O(poly(n))$ number of quantum gates for each $U_i$, they are likely to compose $O(poly(n))$ number of binary coded networks. 
The decomposition of such networks produce, in total, $O(poly(n)M)$ number of quantum gates. On the other had, the amplitude amplification would require $O(\sqrt{M})$ repetitions yielding $O(poly(n)M\sqrt{M})$ number of  CNOT and single quantum gates.  When $M=N$, this becomes $O(poly(n)N\sqrt{N})$.

\subsection{The Circuit Simulation of a Matrix Product}
Let $\mathcal{U}_1$ and $\mathcal{U}_2$ are the universal circuits for the unitary matrices, respectively, $W_1$ and $W_2$.  Then, the circuit for the product $W_1W_2$ can be implemented as $Q_2^{k}\mathcal{U}_2Q_1^{k}\mathcal{U}_1$, where $\mathcal{U}_j$ represents the universal circuit, $Q_j$ is in the form of Eq.\eqref{EqQ1} and $k=\left \lfloor{\frac{\pi}{4}\sqrt{M}} \right \rfloor$. This can be generalized to the product consisting of $r$ number of matrices, $\prod_{j=1}^{r}W_j$ by the following:
\begin{equation}
\label{EqGeneralProduct}
\prod_{j=1}^{r}{ Q_j^{k}\mathcal{U}_j}.
\end{equation}
Here, if each $U_j$ requires $O(N^2)$ quantum gates, the obtained circuit implementation for the matrix product requires $O(r\sqrt{M}N^2)$ quantum gates or $O(\sqrt{M}N^2)$ for $r<<N$. When $U_j$ can be efficiently implementable, e.g. $O(poly(n))$, the matrix product requires  only $O(poly(n)M\sqrt{M})$ number of quantum gates.
 
  \subsubsection{Non-Unitary Matrix Product}
  When the standard amplitude amplification is employed; the amplitude amplification in the second circuit requires the whole amplitude amplification process in the first circuit to be applied again in every iteration. Because of this input dependence, this results in a recursion which makes the whole thing very inefficient.
  
  However, when $U$ is non-unitary but close to a unitary, one can apply the oblivious amplitude amplification. In that case, the complexity of the matrix product becomes the same as the unitary case.

\subsection{ Quantum Circuits for the Functions of a Matrix}
Many trigonometric functions can be represented in the forms of infinite products (Please see the first chapter of Ref.\cite{Zwillinger2014BookSeries}). Some of the infinite product formulas are also valid for matrices. For instance, formulas for the  exponential  and the cosines are as follows:
 Let $A\in C^{\otimes n}$, then the product formula for the exponential of this matrix is defined as \cite{Bernstein2009matrix}:
 \begin{equation}
 \label{EqExpFunc}
e^{A} = \lim_{k \rightarrow \infty}  \left(1+\frac{A}{k}\right)^k
= \lim_{k \rightarrow \infty} W_k^k.
 \end{equation}
 This can be simply proved by the binomial expansion:
 \begin{equation}
 (I+\frac{A}{k})^k=\sum_{j=0}^k \frac{1}{k^j}\frac{k!}{j!(k-j)!}A^j.
 \end{equation}
If we plug this into Eq.(\ref{EqExpFunc}), then we attain the Taylor expansion of the matrix exponential and complete the proof:
  \begin{equation}
\lim_{k\rightarrow \infty}(I+\frac{A}{k})^k
=\lim_{k\rightarrow \infty}\sum_{j=0}^k \frac{1}{k^j}\frac{k!}{j!(k-j)!}A^j=\sum_{j=0}^\infty \frac{1}{j!}A^j=e^A.
 \end{equation}
In a similar fashion, the cosine representation of $A$ can be defined in the product form as:
\begin{equation}
  \begin{split}
cos(\pi A)&=
\prod_{j=1}^{\infty} \left(I-\frac{4A^2}{(2j+1)^2}\right)\\
&= \prod_{j=1}^{\infty} \left(I-\frac{2A}{2j+1}\right)\left(I+\frac{2A}{2j+1}\right) 
=\prod_{j=1}^{\infty} W_{j1}W_{j2}
\end{split}
\end{equation}

As in the matrix product case discussed in the previous section, first each $W_j$ in the product formulas is extended to the form of $U$ given in Eq.\eqref{EqConstructedNonUnitary} where $W_j$ replaces $A$. Then, we generate the circuit designs $\mathcal{U_j}$ for each $W_j$ by either writing $U$ as a sum of simple unitary matrices or following the circuit design in Ref.\cite{Daskin2012universal} used in the numerical examples.  
Then, the same product formula given in Eq.\eqref{EqGeneralProduct} is applied.   
\section{Conclusion}

In this paper, after extending a non-unitary matrix to a unitary matrix, we have  described a method to approximate this unitary extension. 
It is  shown that in most cases the approximation gives a non-unitary matrix sufficiently close to a unitary matrix. Therefore, this matrix can be simulated on an acilla based circuit where
the success probability is increased by the oblivious amplitude amplification.   
Since the oblivious amplitude amplification does not depend on the input to the system register, using the proposed framework one can implement a matrix product for non-unitary matrices by combining two ancilla based quantum circuits. 
This allows quantum computers to  implement the approximation of the matrix functions represented as infinite matrix products. 
A non-unitary matrix can be written in terms of sums of unitary matrices: when each unitary matrix in the summation can be efficiently implementable, then one may also implement the non-unitary  matrix efficiently. Therefore, one may also implement the matrix product and so the approximations of matrix functions efficiently.
We have run numerical simulations for random matrices and shown how the fidelities and the probabilities change in each random case.

%

\end{document}